\pdfoutput=1
\documentclass[aps,prd,reprint,superscriptaddress,longbibliography,nofootinbib]{revtex4-1}

\usepackage{graphicx}
\usepackage{amsmath}
\usepackage{amssymb}
\usepackage{amsfonts}
\usepackage{dcolumn}
\usepackage{dsfont}
\usepackage{latexsym}
\usepackage{rotating}
\usepackage{color}
\usepackage{latexsym}
\usepackage{bbm}
\usepackage{subfigure}
\usepackage{float}
\usepackage{epsfig}
\usepackage{psfrag}
\usepackage{natbib, hyperref}
\usepackage{bm}
\usepackage{amsthm}
\usepackage{eucal}
\usepackage{mathrsfs}
\usepackage{url}
\usepackage{braket}
\usepackage{ulem}

\usepackage{calligra}
\usepackage[T1]{fontenc}
\usepackage[utf8]{inputenc}
\usepackage{newunicodechar}
\usepackage{marvosym}
\usepackage[absolute]{textpos}
\usepackage[cal=cm]{mathalfa}

\usepackage[resetlabels]{multibib}  
\newcites{Supp}{Supplementary~References}   

\usepackage{color} 

\usepackage{hyperref}
\hypersetup{
colorlinks=true,final=true,
        linkcolor=blue,
        citecolor=blue,
        filecolor=blue,
        urlcolor=blue,
}

\begin{document}

\title{Topological Hall Effect and Emergent Skyrmion Crystal\\ in Manganite-Iridate Oxide Interfaces}
\thanks{Copyright notice: This manuscript has been authored by UT-Battelle, LLC under Contract No. DE-AC05-00OR22725 with the U.S. Department of Energy. The United States Government retains and the publisher, by accepting the article for publication, acknowledges that the United States Government retains a non-exclusive, paid-up, irrevocable, world-wide license to publish or reproduce the published form of this manuscript, or allow others to do so, for United States Government purposes. The Department of Energy will provide public access to these results of federally sponsored research in accordance with the DOE Public Access Plan (http://energy.gov/downloads/doe-public-access-plan).}

\author{
Narayan Mohanta$^{1}$,~Elbio Dagotto$^{1,2}$,~Satoshi Okamoto
}

\affiliation{
\textit{Materials Science and Technology Division, Oak Ridge National Laboratory, Oak Ridge, Tennessee 37831, USA}\\
$^{2}$\textit{Department of Physics and Astronomy, The University of Tennessee, Knoxville, TN 37996, USA}\\
}

\begin{abstract}
Scalar spin chirality is expected to induce a finite contribution to the Hall response at low temperatures. We study 
this spin-chirality-driven Hall effect, known as the topological Hall effect, at the manganite side of the interface 
between La$_{1-x}$Sr$_{x}$MnO$_{3}$  and SrIrO$_3$. The ferromagnetic double-exchange hopping at the manganite layer, 
in conjunction with the Dzyaloshinskii-Moriya (DM) interaction which arises at the interface due to broken inversion symmetry and strong 
spin-orbit coupling from the iridate layer, could produce 
a skyrmion-crystal (SkX) phase in the presence of an external magnetic field. Using the Monte Carlo technique 
and a two-orbital spin-fermion model for manganites, supplemented by an in-plane DM interaction, 
we obtain phase diagrams which reveal at low temperatures a clear 
SkX phase and also a low-field spin-spiral phase. Increasing temperature, a skyrmion-gas phase, precursor 
of the SkX phase upon cooling, was identified. The topological Hall effect primarily appears in the SkX phase, 
as observed before in oxide heterostructures. We conclude that the manganite-iridate superlattices 
provide another useful platform to explore a plethora of unconventional magnetic and  transport properties.
 \end{abstract}
           
\maketitle

\section{Introduction}
\label{introduction}
The interplay between spin-orbit coupling and magnetism has led to the emergence of several novel
properties at the interface between distinct transition-metal oxides, such as the anomalous Hall 
effect~\cite{Nagaosa_RMP2009,Nagaosa_PRL2006,HoNyung_NComm2016} and the 
anisotropic magnetoresistance~\cite{Caviglia_PRL2015}. Another source of the Hall effect in 
a heterointerface with both time-reversal and mirror symmetries broken is provided by real-space 
non-collinear magnetic textures with a finite scalar 
spin chirality $\mathbf{S}_i$$\cdot$($\mathbf{S}_j$$\times$$\mathbf{S}_k$)~\cite{Onoda_JPSJ2004,Yi_PRB2009,Hamamoto_PRB2015,Ishizuka_SciAdv2018}. 
This spin chirality generates an effective electromagnetic field for electrons through the
spin Berry phase mechanism. The resulting Hall effect, known as the Topological Hall (TH) effect, 
has been observed in perovskite oxides~\cite{Matsuno_SciAdv2016,Wang_NMat2018,Nakamura_JPSJ2018,Vistoli_NPhys2019}, chiral magnets~\cite{Neubauer_PRL2009,Kanazawa_PRL2011}, frustrated magnets~\cite{Taguchi_Science2001}, and Heusler alloys~\cite{Rana_NJP2016,Swekis_PRM2019,Vir_PRB2019}. 

In compounds with heavy elements and without inversion symmetry, the antisymmetric spin exchange 
such as the Dzyaloshinskii-Moriya (DM) interaction $\mathbf{D}_{ij}$$\cdot$($\mathbf{S}_i$$\times$$\mathbf{S}_j$), 
coexisting with a ferromagnetic (FM) exchange $\mathbf{S}_i$$\cdot$$\mathbf{S}_j$, 
favors a spin-spiral (SS) phase (characterized by a single wave number $\mathbf{q}$)~\cite{Banerjee_NPhys2013,Balents_PRL2014}. 
In the presence of an external magnetic field or an easy-plane anisotropy, the SS competes with ferromagnetism, 
and a skyrmion-crystal (SkX) phase (with three characteristic $\mathbf{q}$ values) emerges at intermediate 
field strengths~\cite{Muhlbauer_Science2009,Ezawa_PRB2011,Huang_PRL2012,Yu_NComm2014}. Another possible route to the SkX phase, 
with higher topological quantum numbers, is to realize a Kondo-type exchange coupling between 
itinerant electrons and localized spins, instead of the DM interaction~\cite{Motome_PRL2017,Motome_PRB2019}.  

An interface between the  ferromagnetic metal La$_{1-x}$Sr$_{x}$MnO$_{3}$ (LSMO) with active $3d$ orbitals 
and  the paramagnetic semimetal SrIrO$_3$ (SIO) with active $5d$ orbitals, is expected to possess a strong DM interaction 
arising from the spin-orbit coupling in SIO and broken structural inversion symmetry at the interface. 
The DM interaction, in bilayers of SIO and SrRuO$_3$, also has been found 
to appear primarily near the interface~\cite{Matsuno_SciAdv2016}. The presence of an interface DM interaction is supported by the 
small lattice mismatch at the interface (lattice constants of SIO and LSMO are $0.394$~nm~\cite{Zhao_JAP2008} and 
$0.388$~nm~\cite{Jalili_ECST2010}, respectively) and a strain-dependent charge transfer~\cite{Okamoto_NanoLett2017}. 
Because of two approximate in-plane mirror symmetries, one mirror plane involves two Mn sites and another reflects two Mn sites, it is reasonable to expect that this DM vector lies in the plane of the interface~\cite{Randeria_PRX2014,Elbio_NSR2019}. Having an in-plane DM vector excludes the possibility of stabilizing 
a conical phase which typically appears in cubic systems such as MnSi~\cite{Buhrandt_PRB2013,Wilson_PRB2014}. On the other hand, 
LSMO, which is well-known to contain several fascinating magnetic phases varying the electronic 
concentration~\cite{Dagotto_PhysRep2001,Adriana_PRL2000,Yunoki_PRL2000,hotta-2003}, 
here is considered to be optimally doped with regards to the 
FM phase. The LSMO-SIO interface is, therefore, a promising platform to search for exotic magnetic textures such as the SkX phase.

In this paper, we investigate the formation of the SkX crystal phase at the LSMO-SIO interface and the influence of 
the non-collinear spin textures on the transverse Hall conductance. We use a spin-fermion model, where itinerant electrons and localized spins are coupled via a Hund interaction, and employ a numerically-exact Monte Carlo (MC) method~\cite{Dagotto_PhysRep2001}. In order to handle large systems, we employ the traveling-cluster approach~\cite{Kumar_EPJB2006,Mukherjee_PRE2015} that allows us to explore the finite-temperature behavior of the SS and the SkX phases. Indeed, one of the main results of the present study is that we observed a SkX phase with N\'eel-type skyrmions 
within a range of magnetic fields at low temperatures.  
Thermal fluctuations give rise to a related phase with spatially disordered nucleated skyrmions, 
a skyrmion gas (SkG), that prevails outside 
of the SkX phase at higher temperatures and acts as a precursor of the SkX phase upon cooling. Also a metastable phase with mixed 
bimerons and skyrmions (BM+Sk)~\cite{Iakovlev_PRB2018} appears at finite temperatures between the SS and SkX phases. 
We construct phase diagrams in the temperature$-$magnetic field plane which describe the parameter regimes where the
Topological Hall effect dominantly appears. 

The strength of the DM interaction depends crucially on several geometrical parameters, such as the thickness 
of the SIO layer~\cite{Matsuno_SciAdv2016}. This offers an external tunability of the interfacial DM interaction and the 
resulting TH effect. The phase diagrams, obtained at different strengths of the DM interaction, thus, serve as a guidance to 
the DM interaction-control of the TH effect. In addition, we also compute the $T$=$0$ phase diagram by comparing 
the total energies of ideal SS, SkX and FM phases and found a reasonable agreement with the low-$T$ properties, 
obtained using the MC method.

The paper is organized as follows: in Sec.~\ref{model},  a lattice model is defined, containing the two-orbital double-exchange model to describe a manganite region, supplemented by the DM interaction and the easy-plane anisotropy. 
We also outline the methodology of MC annealing and computation of physical quantities. In Sec.~\ref{emergence}, we present 
the MC results, revealing the emergence of the SkX phase and its consequences on the transverse Hall conductance. 
In Sec.~\ref{temp_evol}, we show the finite-temperature behavior of 
different phases, obtained using Monte Carlo calculations and discuss the magnetic field vs temperature phase diagrams. 
Section~\ref{T0_phase} describes the $T$=$0$ phase diagram obtained from the total energy calculation 
using ideal spin configurations. In Sec.~\ref{conclusion}, we discuss the relevance of the present results 
to possible experiments on LSMO-SIO films and superlattices, and summarize our results.

\section{Model and Method}
\label{model}
\subsection{Spin-fermion Hamiltonian}
We consider a square lattice which hosts the essential features of LSMO at the two-dimensional interface with SIO \textit{i.e.} the intrinsic magnetism of the manganite layer, supplemented by the induced DM interaction due to the spin-orbit coupling at the iridate layer. 
To describe the hopping of electrons, we use a two-orbital double-exchange Hamiltonian at infinite Hund's coupling. 
As explained, the DM interaction arises from the influence of the iridate layer. Then, the resulting 
spin-fermionic Hamiltonian for the LSMO-SIO interface is given by
\begin{align}
\mathcal{H}\!=\!&-\sum_{\langle ij \rangle}^{\alpha,\beta}\Big(t_{\alpha \beta}^{w} \Omega_{ij}c_{i\alpha}^{\dagger} c_{j\beta} +H.c.\Big)-\sum_{\langle ij \rangle}\mathbf{D}_{ij} \cdot \Big( \mathbf{S}_i \times \mathbf{S}_j  \Big) \nonumber \\
&-h_z\sum_{i}S_{zi}-A\sum_{i}|S_{zi}|^2,
\label{Hamiltonian}
\end{align}
where $c_{i\alpha}^{\dagger}$~\!($c_{i\alpha}$) is the fermionic creation (annihilation) operator at site $i$ with position $\mathbf{r}_i$ and orbital $\alpha$, $t_{\alpha \beta}^{w}$ denotes the nearest-neighbor hopping amplitudes between orbitals $\alpha$ and $\beta$ along the 
hopping direction $w$=$x$,$y$, and is given by $t_{aa}^{x}$=$t_{aa}^{y}$=3$t_{bb}^{x}$=3$t_{bb}^{y}$=$-\sqrt{3}t_{ab}^{x}$=$-\sqrt{3}t_{ba}^{x}$=$\sqrt{3}t_{ab}^{y}$=$\sqrt{3}t_{ba}^{y}$=$3t_0/4$; $a$ and $b$ represent the Mn $e_g$ orbitals, $d_{x^2-y^2}$ and $d_{3z^2-r^2}$, respectively~\cite{Dagotto_PhysRep2001};  the DM vector, acting between nearest-neighbor lattice sites $i$ and $j$, is given by $\mathbf{D}_{ij}$=$D (\mathbf{r}_i -\mathbf{r}_j)$$\times$$\hat z /|\mathbf{r}_i -\mathbf{r}_j|$ with $D$ being the strength of the DM interaction; $h_z$ is the external magnetic field, applied perpendicular to the interface plane; $A$ is the strength of the easy-plane anisotropy, originating from interfacial strain and Rashba spin-orbit coupling~\cite{Randeria_PRX2014}, which is not explicitly included in Eq.\!~\!(\ref{Hamiltonian}); $\Omega_{ij}$ is the effective hopping at infinite Hund's coupling~\cite{Hartmann_PRB1996}, which in terms of the polar and azimuthal spin angles $\theta$ and $\phi$ (assuming the electron spin $\mathbf{S}$ to be described by a classical vector in three-dimensions with amplitude $S$=$3/2$) is given by $\Omega_{ij}$=$\cos{(\theta_i/2)}\cos{(\theta_j/2)}+e^{-i(\phi_i-\phi_j)}\sin{(\theta_i/2)}\sin{(\theta_j/2)}$. We use a square lattice of dimension $N_x$$\times$$N_y$ and numerically solve Hamiltonian~(\ref{Hamiltonian}) using the MC technique to study the spin configuration and transport properties at different sets of parameters. 
We set the hopping amplitude to $t_0$=$1$ and the anisotropy parameter to $A$=$-0.05$ throughout the MC analysis.

\subsection{Monte Carlo annealing}
To obtain the spin configuration at a particular set of parameters, we start at a high temperature $T$=$5$ and slowly cool 
the system down to a desired $T$. In each annealing session, $1000$ temperature steps were employed and at each $T$, 
$2$$\times$$10^{5}$ number of MC steps were used for the spin-configuration update according to the Metropolis algorithm. In each MC step, the spin angle $\theta$ or $\phi$ was changed randomly to $\theta$$\pm$$\Delta \theta$ 
or $\phi$$\pm$$\Delta \phi$, respectively, where $\Delta \theta$ and $\Delta \phi$ are the step sizes of the angles, 
set to 5 degrees throughout the paper. The diagonalization of the fermionic part of the Hamiltonian is numerically 
expensive and to alleviate this problem, we used the traveling-cluster update scheme~\cite{Kumar_EPJB2006,Mukherjee_PRE2015}, in which a smaller cluster of size $N_{xc}$$\times$$N_{yc}$ is used for the MC spin update. The results presented here were obtained using a $20$$\times$$20$ lattice and $5$$\times$$5$ traveling cluster, with open-boundary conditions. We checked, with a larger lattice of size $32$$\times$$32$ and 
a larger traveling cluster of size $10$$\times$$10$, for any finite-size effects and obtained a consistent description. The observables were calculated from the thermalized spin configurations and were MC averaged, at each $T$ and $h_z$, with $100$ different realizations.

\subsection{Calculation of observables}
\noindent \textit{Skyrmion number.} A magnetic skyrmion has a distinct topological structure and when the underlying lattice is transformed from a torus to a sphere, the skyrmion gives a full coverage of the sphere. This unique feature enables skyrmions to be classified by a topological index, called the skyrmion number, which is expressed as~\cite{Heinze_NPhys2011,Nagaosa_NNat2013}
\begin{align}
N_{sk}=\frac{1}{4\pi} \int \mathbf{S} \cdot \Bigg( \frac{\partial \mathbf{S}}{\partial x} \times \frac{\partial \mathbf{S}}{\partial y} \Bigg) dx dy.
\label{sk_num}
\end{align}
In practice for a square lattice with discrete points, the integration in Eq.~(\ref{sk_num}) has been performed by using a summation over the underlying lattice sites and the partial derivatives have been calculated using a central-difference scheme within the five-point stencil method~\cite{Ding_JCP2006}.\\

\noindent \textit{Spin correlation function.} An important quantity to identify magnetic phases and track thermodynamic phase transitions is the spin correlation function. We take the Fourier transform of the real-space spin-spin correlation function as follows
\begin{align}
 S_{\mathbf{q}} =\frac{1}{N}\sum_{ij}^{|\mathbf{r}_{ij}|
<\delta} \langle \mathbf{S}_{i} \cdot \mathbf{S}_{j} \rangle e^{-i\mathbf{q} \cdot \mathbf{r}_{ij} },
\label{Sq}
\end{align}
where $\mathbf{S}_{i} $ denotes the localized spin at site $i$,  $\mathbf{r}_{ij}$ 
is the distance between sites $i$ and $j$, $\delta$ is the radius of a circle around site $i$ within which all sites 
are considered to calculate the correlation function, and $N$ is the total number of lattice sites. We use the radius 
$\delta$ up to $N_x/2$. If some sites $j$ landed outside the full cluster, they were discarded. The quantity in Eq.~(\ref{Sq}) can be compared to intensity measured in neutron-scattering experiments and, therefore, we use the Bragg intensity as $I(\mathbf{q})$=$ S_\mathbf{q}$.\\

\noindent \textit{Hall conductance.} The transverse Hall conductance is obtained by the current-current correlation function, described by the Kubo formula~\cite{Yi_PRB2009}, as given below
\begin{align}
\sigma_{xy}\!=\!\frac{e^2}{h}\frac{2\pi}{N} \! \sum_{\epsilon_m \neq \epsilon_n} \! \frac{f_m-f_n}{(\epsilon_m-\epsilon_n)^2+\eta^2} \text{Im.}\Big( \langle m | \hat{j_x} | n \rangle \langle n | \hat{j_y} | m \rangle \Big),
\end{align}
where $f_m$ is the Fermi function at temperature $T$ and energy $\epsilon_m$, $\hat{j}_x$=$i\sum_{i\alpha\beta}(t_{\alpha \beta}^{x} \Omega_{i,i+\hat{x}}c_{i\alpha}^{\dagger} c_{i+\hat{x}\beta}-t_{\alpha \beta}^{x*} \Omega_{i,i+\hat{x}}^{*}c_{i+\hat{x}\beta}^{\dagger} c_{i\alpha})$ is the current operator along the $x$ direction, and $\eta$ is the relaxation rate which is kept fixed at a value $0.1$.
\begin{figure}[t]
\begin{minipage}[c]{0.47\textwidth}
\centering
\epsfig{file=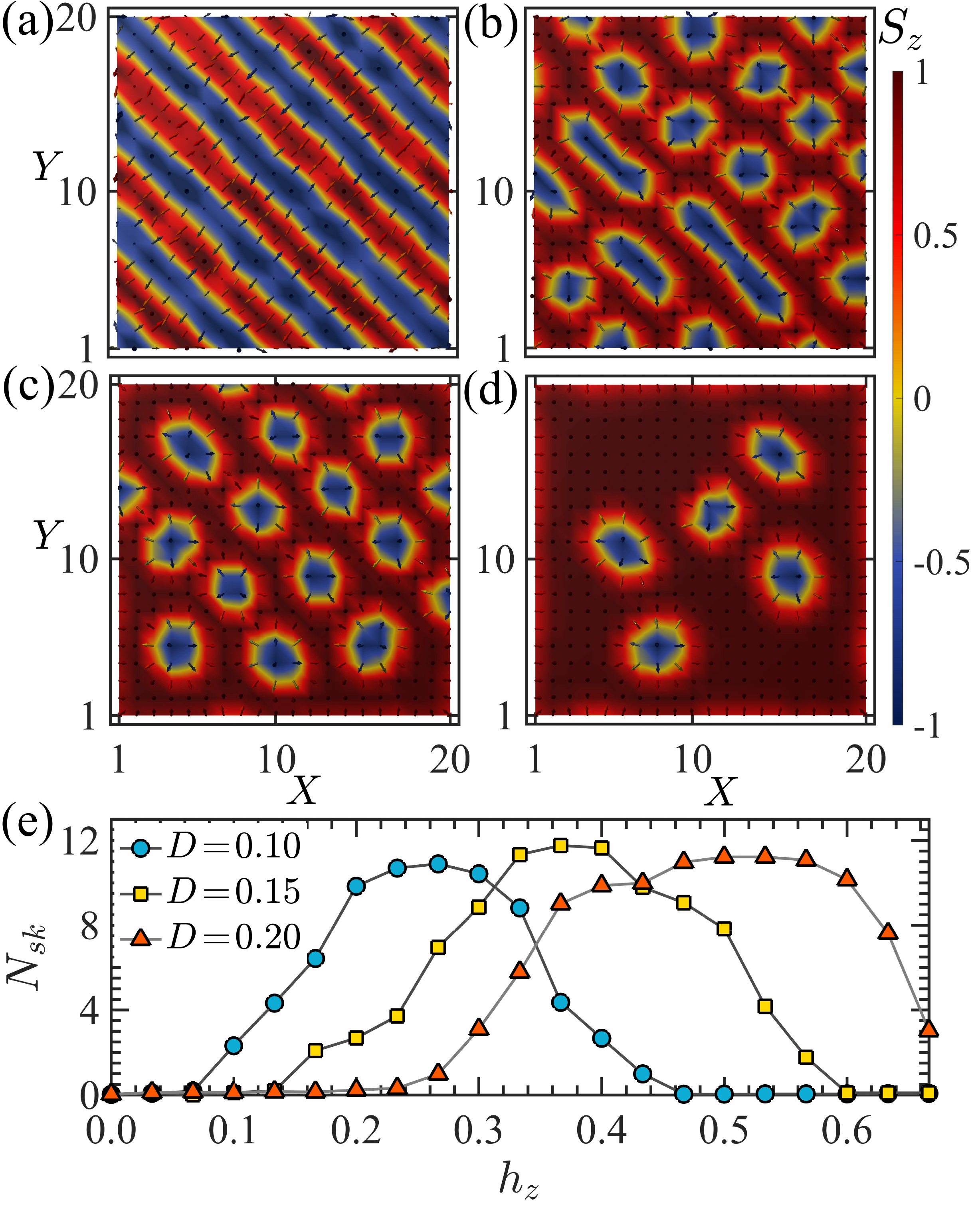,trim=0.0in 0.0in 0.0in 0.0in,clip=false, width=85mm}
\caption{Typical real-space spin configurations (a)-(d) obtained during the Monte Carlo time evolution on a $20$$\times$$20$ cluster with open boundary conditions, revealing spin-spiral (SS), mixed bimeron+skyrmion (BM+Sk), skyrmion crystal (SkX) and skyrmion gas (SkG) phases at different magnetic fields ($h_z$) and temperatures ($T$) given by (a) $h_z$=$0$, $T$=$0.001$, (b) $h_z$=$0.2$, $T$=$0.001$, (c) $h_z$=$0.27$, $T$=$0.001$, and (d) $h_z$=$0.27$, $T$=$0.15$, respectively, at a fixed value of the DM interaction $D$=$0.1$. The arrows denote the in-plane components ($\mathbf{S}_{xi}$,$\mathbf{S}_{yi}$) while the colorbar represents the normalized perpendicular component of magnetization $\mathbf{S}_{zi}$. (e) Variation of the skyrmion number $N_{sk}$ as a function of $h_z$, at the values of $D$. The results show a clear enhancement in $N_{sk}$ within a range of field values. The hopping amplitude is fixed to $t_0$=$1$ and the anisotropy strength to $A$=$-0.05$.}
\label{fig1}
\end{minipage}
\end{figure}
\begin{figure}[t]
\centering
\begin{minipage}[c]{0.47\textwidth}
\centering
\epsfig{file=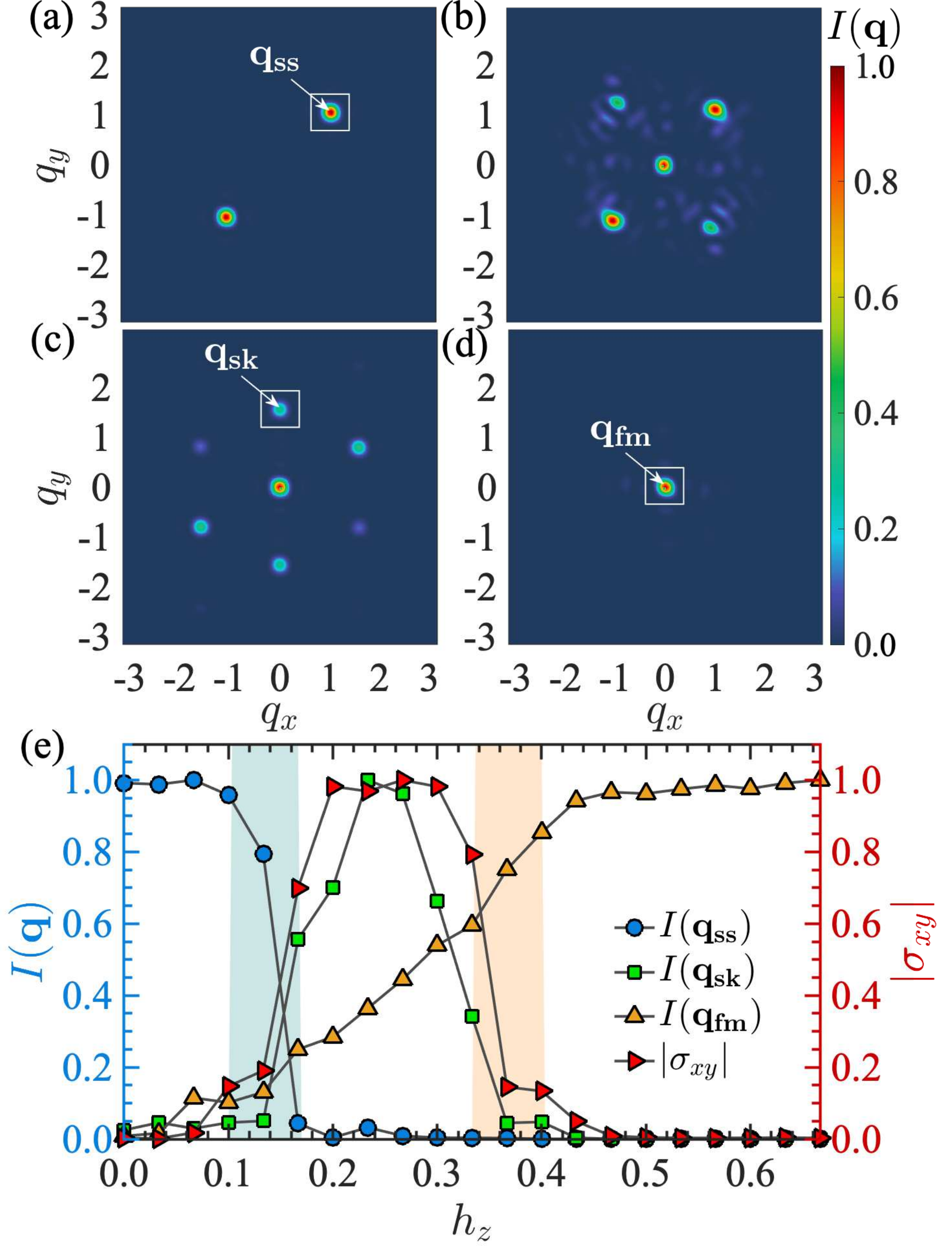,trim=0.0in 0.0in 0.0in 0.0in,clip=false, width=85mm}
\caption{Monte Carlo averaged intensity profile of the Bragg intensity $I(\mathbf{q}$) for magnetic field values (a) $h_z$=$0$, (b) $h_z$=$0.2$, (c) $h_z$=$0.27$, and (d) $h_z$=$0.37$, respectively. (e)  Variation of the 
normalized $I(\mathbf{q}$) (left y axis) and  Hall conductance $\sigma_{xy}$ (right y axis) with $h_z$, at different characteristic 
momenta $\mathbf{q}$=$\mathbf{q_{ss}}$, $\mathbf{q}$=$\mathbf{q_{sk}}$, $\mathbf{q}$=$\mathbf{q_{fm}}$, as indicated, respectively, 
in panels (a),~(c) and (d).  Panel (b) corresponds to the narrow bimeron regime, typically containing segments of ``stripes'' (panel (a)) 
involving both diagonals in the same configuration. The blue and yellow  ranges of $h_z$ in panel (e) show the field ranges within which the phase transitions (which are likely first order in nature at $T$=$0$) take place. In all panels, the DM interaction is $D$=$0.1$, the hopping amplitude $t_0$=$1$, 
the anisotropy strength $A$=$-0.05$, and the temperature $T$=$0.001$.}
\label{fig2}
\end{minipage}
\end{figure}
\section{Emergence of skyrmion crystal}
\label{emergence}
We started our effort by confirming the well-known fact that the double-exchange hopping term in the absence of the DM interaction 
and magnetic field, yields a FM phase at low temperatures $T$$\lesssim$$0.2$~\cite{yunoki-2000}. Then, after incorporating a finite DM interaction, 
a single-$\mathbf{q}$ SS phase appears at low temperatures, as shown in Fig.~\ref{fig1}(a) at $T$=$0.001$. With increasing the DM 
strengh $D$, the period of the spiral decreases gradually. There are two degenerate, diagonally-opposite spiral solutions which often 
merge together to form a labyrinth-like metastable spin configuration. This type of metastable spin configuration can be avoided 
by re-annealing the spin configuration obtained at low $T$ from the previous MC annealing process. To perform the re-annealing process, 
we take the previous annealed spin configuration to a temperature $T$=$0.5$ and slowly cool down to the lowest temperature $T$=$0.001$. 
We considered 20 re-annealing sessions to obtain a stable spin configuration. 
\begin{figure}[t]
\centering
\epsfig{file=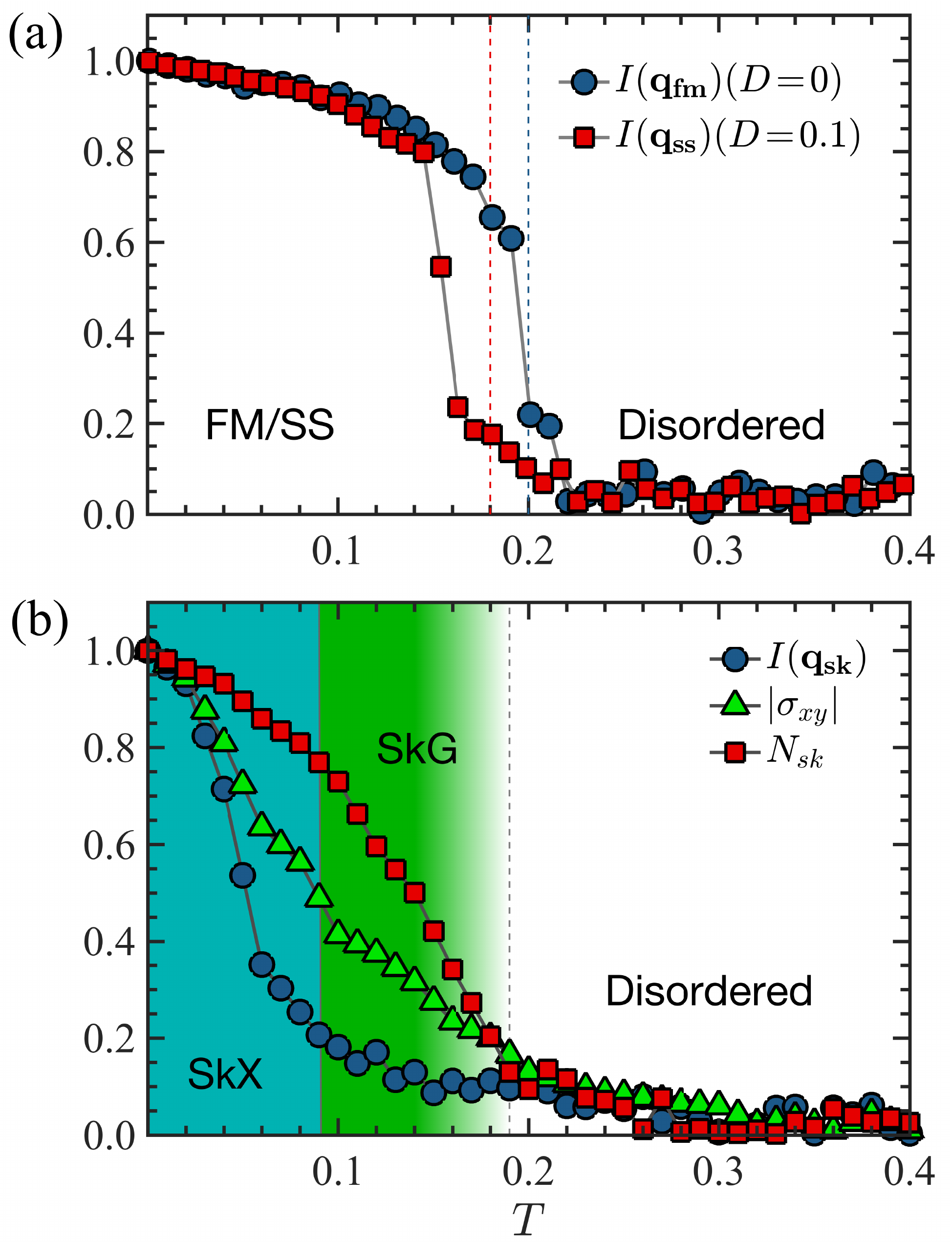,trim=0.0in 0.0in 0.0in 0.0in,clip=false, width=85mm}
\caption{(a) Monte Carlo results showing the temperature variation of the intensity $I(\mathbf{q_{fm}}$) for the FM phase 
(with zero DM interaction strength $D$=$0$, nonzero anisotropy constant $A$=$-0.05$, 
and zero magnetic field $h_z$=$0$) and intensity $I(\mathbf{q_{ss}}$) for the spin-spiral (SS) phase 
(at $D$=$0.1$ and $h_z$=$0$). The blue and red dashed lines provide a rough estimation of
the critical temperatures here crudely defined as the temperatures where $I(\mathbf{q_{fm}}$) 
and $I(\mathbf{q_{ss}}$) drop to 20$\%$ of the $T$=$0.001$ value. Panel (b) are also Monte Carlo 
results for the intensity $I(\mathbf{q_{sk}}$) in the SkX regime, 
Hall conductance $|\sigma_{xy}|$, and skyrmion number $N_{sk}$ (at $D$=$0.1$ and $h_z$=$0.27$). The critical temperature where 
$I(\mathbf{q_{sk}}$ drops to 20$\%$ of the value at $T$=$0.001$ roughly defines the phase boundary of the skyrmion crystal (SkX) phase. 
Similarly, $N_{sk}$ defines the phase boundary of the skyrmion gas (SkG) phase. All quantities in panels (a) and (b) are normalized 
to the respective values at the lowest temperature $T$=$0.001$. Other parameters ($t_0$ and $A$) are the same as in Fig.~1. 
Using other reasonable cutoffs criteria to define critical temperatures leads to similar phase diagrams.}
\label{fig3}
\end{figure}

Following this multiple-annealing process, in the presence of an external magnetic field we observed that, beyond a critical field $h_z$,
the SS phase is transformed into a SkX phase where the skyrmions arrange themselves in a nearly-triangular crystal, as depicted 
in Fig.~\ref{fig1}(c) at $h_z$=$0.27$ and $T$=$0.001$. Thermal fluctuations or disorder introduce a metastable regime, 
a mixed bimeron+skyrmion (BM+Sk) phase, between the SS and the SkX phase, as shown in Fig.~\ref{fig1}(b) at $h_z$=$0.14$ and $T$=$0.05$. 
The bimerons are extended skyrmions and have finite contributions to the scalar spin chirality, similar to the skyrmions. 
With a further increase in $h_z$, the SkX phase melts into the SkG phase which is a gas of nucleated skyrmions, and finally transforms 
into a fully-polarized FM phase. The SkG phase appears within a very narrow range of $h_z$ at low temperatures and further
work is needed to confirm its presence at those temperatures. But SkG is robust and primarily dominates at higher temperatures. For example, 
a typical spin configuration of the SkG phase, obtained at $h_z$=$0.27$ and $T$=$0.15$, is shown in Fig.~\ref{fig1}(d). 

The variation of the skyrmion number $N_{sk}$ with $h_z$ is depicted in Fig.~\ref{fig1}(e) for three 
different values of $D$, revealing a clear enhancement within 
a range of $h_z$. At $T$=$0$, $N_{sk}$ should exhibit sharp first-order transitions at two critical values of $h_z$, clearly distinguishing 
the SkX phase from others. But here due to thermal fluctuations, the expected sharp first-order jumps are replaced by 
smooth crossovers as in Fig.~\ref{fig1}(e). Increasing $D$, the $h_z$ range where the SkX phase appears is enhanced. 
This is an anticipated behavior, since a stronger DM interaction helps to stabilize the SkX phase.

In Fig.~\ref{fig2}(a)-(d), we plot the MC-averaged Bragg intensity profile for the four different phases discussed in Fig.~\ref{fig1}, \textit{viz.} the SS, BM+Sk, SkX, and SkG. The SS and the SkX phases show, respectively, the single-$\mathbf{q}$ (Fig.~\ref{fig2}(a)) 
and triple-$\mathbf{q}$ (Fig.~\ref{fig2}(c)) structures of the spin configuration. Figure~\ref{fig2}(b), for the BM+Sk phase, interestingly, displays 
a double-$\mathbf{q}$ spin configuration which is absent at $T$=$0$. Because of the large FM fraction, the SkG phase is dominated by the Bragg intensity at $\mathbf{q}$=$\mathbf{0}$ but with non-zero intensities at $\mathbf{q}$$\ne$$\mathbf{0}$ due to the randomly distributed skyrmions. Thus, $I(\mathbf{q_{fm}})$ in the SkG phase is not fully developed while $I(\mathbf{q_{sk}})$ is suppressed.  The Bragg intensity $I(\mathbf{q})$ at different characteristic momenta, reveal a sequence of phase transitions, as shown in Fig.~\ref{fig2}(e) 
where the transverse Hall conductance $\sigma_{xy}$ is also plotted. Evidently, $\sigma_{xy}$ (and also $N_{sk}$ in Fig.~\ref{fig1}(e)) become 
enhanced within a broader range of $h_z$ than $I(\mathbf{q_{sk}})$, where $\mathbf{q}$=$\mathbf{q_{sk}}$ is the characteristic momentum for 
the SkX phase (as defined in Fig.~\ref{fig2}(c)). The SS and SkX phases can, therefore, be identified using $I(\mathbf{q_{ss}})$ and $I(\mathbf{q_{sk}})$, respectively, whereas, $\sigma_{xy}$ and $N_{sk}$ are suitable to extract the BM+Sk and the SkG phases.

\begin{figure}[t]
\centering
\epsfig{file=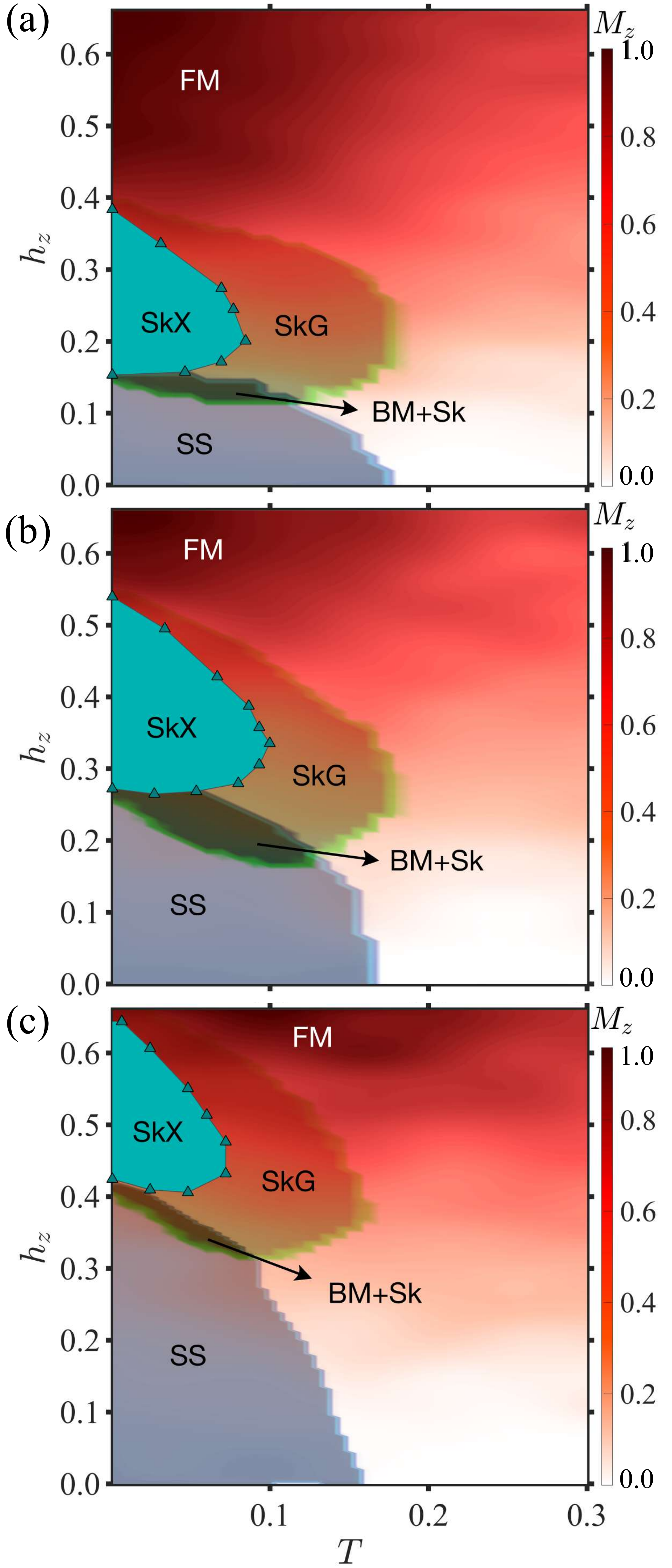,trim=0.0in 0.0in 0.0in 0.0in,clip=false, width=75mm}
\caption{Temperature ($T$) vs. magnetic field ($h_z$) phase diagrams for three different values of the DM interaction (a) $D$=$0.1$, (b) $D$=$0.15$ and (c) $D$=$0.2$. The colorbar represents the normalized average magnetization, perpendicular to the interface plane. The hopping amplitude is fixed to $t_0$=$1$ and the easy-plane anisotropy strength to $A$=$-0.05$.}
\label{fig4}
\end{figure}

\section{Temperature evolution}
\label{temp_evol} 

Having identified the phases appearing at various values of the external magnetic field at low temperatures, we explore the finite-temperature behavior of these phases. In Fig.~\ref{fig3}(a), we show the temperature variation 
of the Bragg intensities $I(\mathbf{q_{fm}})$ and $I(\mathbf{q_{ss}})$ for the FM and SS phases in the absence of 
external magnetic fields at $D$=$0$ and $D$=$0.1$, respectively. The critical temperature $T_c^{SS}$$\simeq$$0.18$ for 
the SS phase appears to be rather close to that of the FM phase ($T_c^{FM}$$\simeq$$0.2$) 
which exists at the two-dimensional interface in the absence of DM interaction and external magnetic fields. 
For the phases associated with the skyrmions, we plot the Bragg intensity $I(\mathbf{q_{ss}})$, Hall conductance $\sigma_{xy}$, 
and skyrmion number $N_{sk}$ vs. $T$ in Fig.~\ref{fig3}(b) at a field $h_z$=$0.27$. Evidently, $I(\mathbf{q_{sk}})$ drops 
faster with $T$ than the other two observables, indicating that the SkX phase (here determined by an 80$\%$ drop 
in $I(\mathbf{q_{sk}})$, but other conventions lead to similar conclusions) exists at much lower temperatures than 
the SkG phase (determined by an 80$\%$ drop in $N_{sk}$). The critical temperatures for the SkX and the SkG phases at $h_z$=$0.27$ are roughly $T_c^{SkX}$$\simeq$$0.09$ and $T_c^{SkG}$$\simeq$$0.19$, respectively. The TH effect, although 
strongest at the SkX phase, exists at temperatures much above the SkX phase. There could be other contributions to the TH effect, and 
one potential origin is the skew scattering induced by the scalar spin chirality~\cite{Ishizuka_SciAdv2018}. The skew-scattering 
induced TH effect appears near the transition to the SkX phase and is reflected by a change in the sign of the Hall conductance. 
In addition to the SkG phase, the BM+Sk phase also yields a finite contribution to the TH effect, as we shall discuss below.

We constructed the phase diagrams for the manganite-iridate single interface in the temperature vs. magnetic field plane. Results
are depicted in Fig.~\ref{fig4} for three different values of the DM strength $D$. The phase diagrams show 
five different phases already discussed before \textit{viz.} SS, SkX, BM+Sk, SkG and FM. Note the BM+Sk and SkG regions
are suppressed at very low temperatures. The FM phase, identified by both the Bragg intensity $I(\mathbf{q_{fm}})$ and 
the average out-of-plane magnetization $M_z$, does not have any phase boundary in the $h_z$-$T$ plane and prevails 
in the high-field regime. The BM+Sk phase is identified by the overlap of the SS and the SkG phases, obtained, respectively, 
by $I(\mathbf{q_{ss}})$ and $N_{sk}$.

Increasing $D$, the SS phase expands towards larger fields, which is expected since the DM interaction helps in stabilizing 
this SS phase. However, the critical temperature $T_c^{SS}$ tends to decrease at very large $D$ (noticeable in Fig.~\ref{fig4}(c) 
for $D$=$0.2$). Very large values of $D$ results in a SS phase with a spiral of short wavelength which makes the SS phase 
less susceptible to thermal fluctuations, accounting for the slow decrease in $T_c^{SS}$ with increasing $D$. 
The SkX phase also moves towards higher-fields with increasing $D$ and is the largest in size (among the three 
plots shown in Fig.~\ref{fig4}) for $D$=$0.15$. For very large $D$, the skyrmion sizes are much smaller, comparable 
to the square lattice spacing, and, thus, the resulting SkX phase is vulnerable to thermal fluctuations and disorder. 
The results suggest that both the SS and the SkX phases are strongest at an optimal value of the DM interaction strength. 

The results presented thus far are for a fixed value of the easy-plane anisotropy parameter $A$=$-0.05$. We have  verified 
that a low-to-moderate value of the easy-plane anisotropy stabilizes the SkX phase, as found in a previous study~\cite{Randeria_PRX2014}.

\section{$\mathbf{T=0}$ Phase diagram}
\label{T0_phase}
\begin{figure}[t]
\centering
\epsfig{file=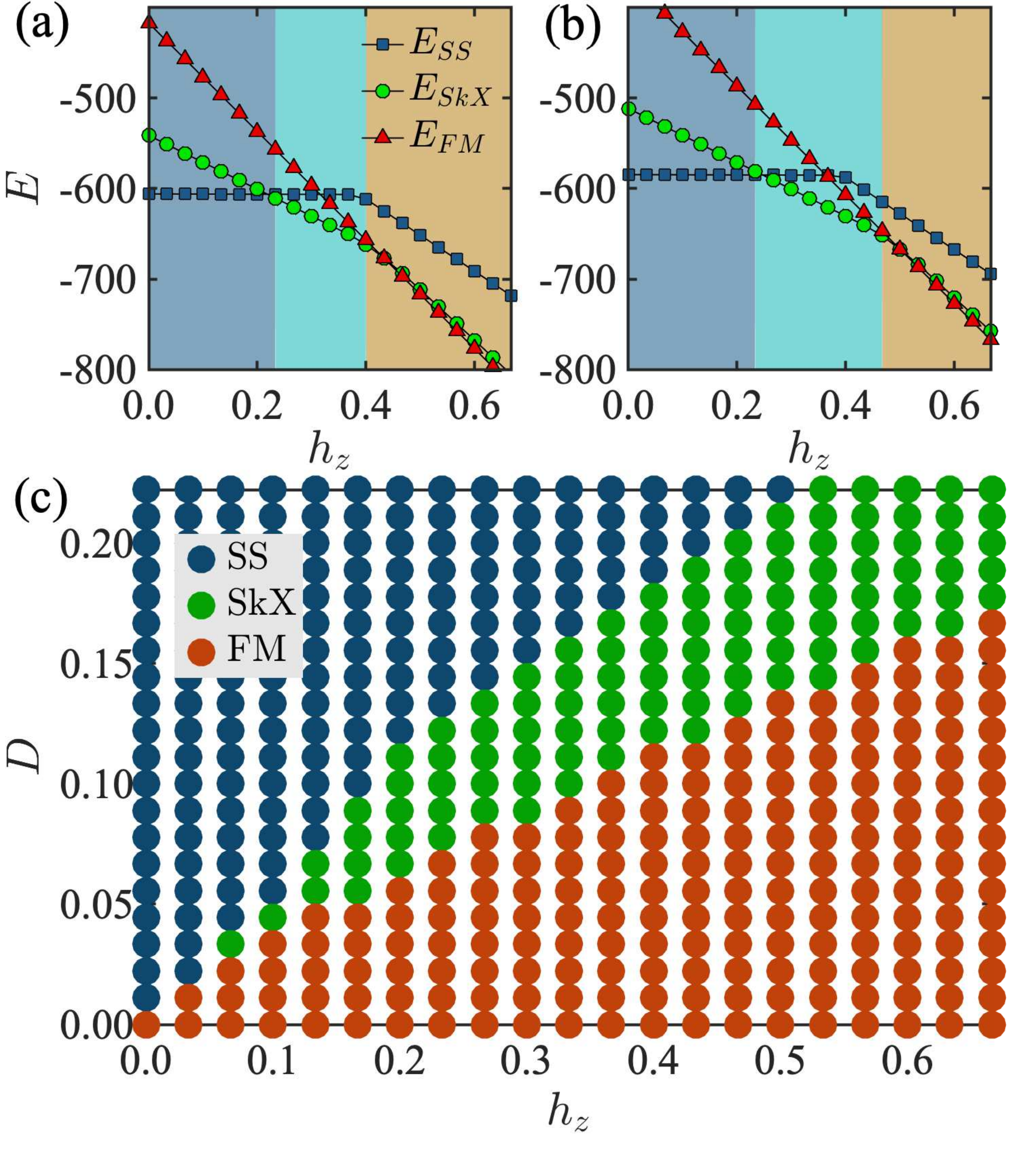,trim=0.0in 0.0in 0.0in 0.0in,clip=false, width=88mm}
\caption{Total energy of the  ideal spin-spiral (SS), skyrmion-crystal (SkX) and ferromagnetic (FM) phases 
at zero temperature ($T$=$0$) vs. the strength of the magnetic field $h_z$, at a fixed DM interaction strength $D$=$0.1$. 
Panel (a) is at easy-plane anisotropy strength $A$=$-0.05$ and (b) at $A$=$-0.1$. These two panels show 
that the minimum-energy condition is satisfied for the SkX phase within a range of $h_z$ values, in agreement with MC results. 
(c) Zero-temperature phase diagram in the $h_z$-$D$ plane for $A$=$-0.05$ and $t_0$=$1$.}
\label{fig5}
\end{figure}
In the MC analysis discussed above, the lowest temperature accessed was $T$=$0.001$. Below this temperature the MC procedure 
is less accurate because it is difficult to evolve away from nearly-frozen metastable states. To complement the MC analysis with
zero temperature results, 
we performed calculations of total energies of the three phases SS, SkX, and FM, by considering ideal spin configurations. We consider 
a $20$$\times$$20$ lattice with open boundary conditions, same as in above MC analysis, 
to compute the total energies of the three phases using the Hamiltonian~Eq.(\ref{Hamiltonian}), after 
optimizing the period of the spiral in the SS phase and the skyrmion radius and skyrmion-skyrmion separation in the SkX phase. 

The variation of the total energies $E_{SS}$, $E_{SkX}$ and $E_{FM}$ for the SS, SkX and the FM phases, respectively, with respect 
to $h_z$ at fixed DM interaction $D$=$0.1$ is shown in Fig.~\ref{fig5}, for two different values of the anisotropy 
parameter $A$=$-0.05$ and $A$=$-0.1$. By the minimum energy criterion, we can identify the most favorable spin configurations. 
For $A$=$-0.05$, we find that in the low-field regime $h_z$$\lesssim$$0.23$, the SS phase has the lowest energy, while for 
$h_z$$\geq$$0.4$, the FM phase has the lowest energy. Remarkably, the SkX phase wins within an intermediate field range 
$0.23$$<$$h_z$$<$$0.4$, which is close to the field range predicted by the MC simulations. The transitions from the SS phase 
to the SkX phase and that from the SkX phase to the FM phase are first order at $T$=$0$ (level crossing). 
With $A$=$-0.1$, we clearly observe that the SkX phase becomes wider with regards to the field range. 
It is interesting to note that $E_{SS}$ is quite insensitive to the change in $h_z$ in the SS phase and changes slope 
towards the second critical field value near the boundary between the SkX and the FM phases. $E_{SkX}$ also 
changes its slope at the transition to the FM phase, because of the drastic enhancement of the optimal radius of the skyrmions beyond a critical $h_z$, as noted before~\cite{Randeria_PRX2014}. We tune $D$ and plot the $T$=$0$ phase diagram in 
Fig.~\ref{fig5}(c) which reveals that, in general, the SkX phase becomes stronger with higher DM interaction and higher magnetic field.

\section{Discussion and Conclusion}
\label{conclusion}
The experimental critical temperature for the FM phase of a manganite-iridate interface 
appears within a range $100 {\rm K} \alt T_c^{FM} \alt 300 {\rm K}$, depending upon 
the thickness of the manganite layer~\cite{HoNyung_NComm2016}. From the MC analysis, 
we find $T_c^{FM}$$\approx$$0.2t_0$ (with anisotropy parameter $A$=$-0.05$), which gives our hopping energy scale $t_0$(=$T_c^{FM}/0.2$) as $40 {\rm meV} \alt t_0 \alt 130 {\rm meV}$.  For a purely classical spin model with Heisenberg exchange term $-J\sum_{\langle ij \rangle}\mathbf{S}_i$$\cdot$$\mathbf{S}_j$, instead of the double-exchange term, we obtain $T_c^{FM}$$\simeq$$7.5J$ with an anisotropy parameter $A$=$-0.05J$ (results gathered using a 20$\times$20 cluster and MC simulations, not shown). 
By comparing the critical temperatures for the FM phase, obtained separately from the double-exchange model and 
classical spin model, we find the effective Heisenberg-exchange parameter for the considered double-exchange model to be $J$$\approx$$0.03t_0$. The easy-plane anisotropy for LSMO films has been found to be $AS^2$$\approx$$0.21$~meV~\cite{Boschker_JMMM2011} which gives $A$$\approx$$0.09$~meV with $S$=$3/2$. Therefore, the critical DM interaction strength to realize 
the skyrmion crystal $D_c$$\sim$$\sqrt{JA}$~\cite{Herve_NComm2018,Rossler_Nature2006,Rohart_PRB2013}  
appears in a range $0.3 {\rm meV} \alt  D_c \alt {\rm 0.6}$~meV. LSMO also has a weak antiferromagnetic coupling~\cite{Dagotto_PhysRep2001} 
which has not been included in our description. Such a competing antiferromagnetic coupling will reduce the 
MC-estimated $T_c^{FM}$ and, in turn, enhance $D_c$.  

We have realized skyrmions of radius approximately $3$ unit cells with $D$=$0.1t_0$. Increasing $D$ further 
will reduce the size of the skyrmions and when $D$ is comparable to the hopping energy scale $t_0$, the skyrmion 
size will become too small to be studied within the MC method. Conversely, very small values of $D$ make the skyrmion 
size very large compared to the lattice size. We, therefore, have considered intermediate values of $D$ for 
which the skyrmion size is optimal for studying the SkX phase in the finite clusters where the spin-fermion model can
be studied numerically. For similar reasons, the critical magnetic fields, obtained from the MC analysis, 
are unrealistically large. The main purpose here is to provide the qualitative variations of the skyrmion phases with D. In reality, we anticipate  similar behavior in smaller energy scales. 


To summarize, we have investigated the Topological Hall effect arising from the scalar spin chirality 
of an emergent skyrmion crystal using a spin-fermionic model for a manganite-iridate interface. 
Using Monte Carlo calculations, we realized a nearly-triangular crystal of N\'eel-type skyrmions, arising within a finite range 
of external magnetic fields. A gas phase of well-formed independent skyrmions was also observed 
primarily at higher temperatures above 
the skyrmion crystal phase. Also, a mixed bimeron+skyrmion phase appears at finite temperatures between the spin-spiral 
phase and the skyrmion crystal phase. Topological Hall measurements, together with neutron-scattering experiments, 
can detect these complex phases.

We conclude that manganite-iridate interfaces offer a unique platform to explore unconventional magnetic and transport properties. 
We have focused on the doping range in which LSMO is in its FM phase. Proximity effect from other types of magnetic 
phases, such as the antiferromagnetic phase or CE states, along with the large DM interaction, 
can lead to interesting phenomena~\cite{Dong_PRL2009}. The easy-plane anisotropy, which stabilizes the skyrmion 
crystal phase, can be modified using epitaxial strain, providing a useful knob to tune the TH effect. 
The DM interaction can be controlled by changing the thickness of the iridate layer, while the 
manganite layer thickness governs the level of spin polarization at the interface, opening a 
novel path towards efficient control of the TH effect by engineering multi-layer heterostructures. \\

\section*{acknowledgments}
The authors gratefully acknowledge John Nichols, Elizabeth M. Skoropata, and Ho Nyung Lee  for discussions on their related experiments. All members of this collaboration were supported by the U.S. Department of Energy (DOE), Office of Science, Basic Energy Sciences (BES), Materials Sciences and Engineering Division. This research used resources of the Compute and Data Environment for Science (CADES) at the Oak Ridge National Laboratory, which is supported by the Office of Science of the U.S. Department of Energy under Contract No. DE-AC05-00OR22725.


%

\end{document}